\documentclass[twocolumn,showpacs,preprintnumbers,amsmath,amssymb]{revtex4}

\usepackage{graphicx}
\usepackage{bm}

\begin{document}
\title{
Scaling laws for the photo-ionisation cross section of two-electron atoms
}

\author{Chang Woo Byun}
\author{Nark Nyul Choi}
\author{Min-Ho Lee}
\affiliation{School of Natural Science, Kumoh National Institute of Technology,
Kumi, Kyungbook 730-701, Korea.}
\author{Gregor Tanner}
\affiliation{School of Mathematical Sciences,
University of Nottingham, University Park, Nottingham NG7 2RD, UK.}

\date{\today}

\begin{abstract}
The cross sections for single-electron photo-ionisation in two-electron atoms 
show fluctuations which
decrease in amplitude when approaching the double-ionisation threshold. Based
on semiclassical closed orbit theory, we show that the algebraic decay of the
fluctuations can be characterised in terms of a threshold law
$\sigma \propto |E|^{\mu}$ as $E \to 0_-$ with exponent $\mu$ 
obtained 
as a combination of stability exponents of the triple-collision
singularity. It differs from Wannier's exponent dominating double
ionisation processes. The details of the fluctuations are linked to a set of
infinitely unstable classical orbits starting and ending in the
non-regularisable triple collision. The findings are compared with quantum
calculations for a model system, namely collinear helium.
\end{abstract}

\maketitle
 
Recent experimental progress has significantly improved the energy
resolutions of highly excited two electron states
below \cite{Pue01,Cza04} and above \cite{double} the
double ionisation threshold $E=0$. Near the threshold, electron-electron
correlation effects become dominant which are directly observable in
total and partial cross sections, see \cite{KA00,Sad00,TRR00} for 
recent reviews; an example is Wannier's celebrated threshold law 
for double ionisation \cite{Wa53} confirmed experimentally in \cite{KSA88}. 
For $E<0$, two-electron atoms exhibit a rich resonance 
spectrum while the classical dynamics of this three-body Coulomb problem 
becomes chaotic. Approaching the double ionisation threshold from 
below has thus proved challenging \cite{TRR00} and recent experimental 
and theoretical efforts still only reach single-ionisation
thresholds $I_N$ with $N \le 15$  \cite{Pue01,Cza04,Bue95}.
Semiclassical methods need to address the chaotic 
nature of the classical dynamics which is dominated by the complex folding 
patterns of the stable/unstable manifolds of the triple collision \cite{CLT04}.
Due to the high dimensionality of the system semiclassical applications have 
been restricted to subsets of the full spectrum and again small 
$N$ values \cite{TRR00}. 

In this letter, we show that the fluctuations in the total 
cross section for single electron photo-ionisation below the 
three-particle breakup energy
decays algebraically with an exponent determined by the triple collision 
singularity {\em different} from Wannier's exponent. Writing the
cross section in dipole approximation \cite{Bog88, DD88}, we obtain
\begin{equation} \label{cs-G}
\sigma(E) = -4 \pi \, \alpha \, \hbar \omega \,
\Im \langle D \phi_i |G(E) |D \phi_i\rangle
\end{equation}
where $\phi_i$ is the wave function of the initial state
and $D = {\bm \pi} \cdot ({\bf r}_1 + {\bf r}_2)$ is the dipole operator with
$\bf \pi$,  the polarisation of the incoming photon and ${\bf r}_j$, the
position of electron $j$. Furthermore, $G(E)$ is the Green function of the
system at energy $E = E_i + \hbar \omega$ and $\alpha = e^2/\hbar c$.
Note that we work in the infinite nucleus mass 
approximation, that is, the position of the nucleus is fixed at 
the origin.

By expressing the Green function semiclassically in terms of classical 
trajectories \cite{Bog88}, fluctuations in the cross section 
of hydrogen-like atoms in external fields have been analysed 
successfully using {\em closed orbit theory} (COT) \cite{Bog88,DD88}. 
In the semiclassical limit, the support of the wave function $\phi_i$
shrinks to zero relative to the size of the system reducing the
integration in (\ref{cs-G}) to an evaluation of the Green function at the
origin. This is strictly valid only for potentials sufficiently smooth 
at the origin, corrections due to diffractive scattering at the
central singularity give additional contributions often
treated in quantum defect approximation \cite{Del94,QD}.
The situation is different for two-electron atoms where the dynamics
near the origin is dominated by the non-regularisable triple collision
singularity. Closed orbit theory has been used to analyse experimental
photoabsorbtion spectra of helium with and without
external fields \cite{Del94,He} for highly asymmetric states;
accompanying theoretical considerations treat the system in a single
electron approximation thus not considering the triple collision dynamics
which becomes important for doubly-excited resonances.

In the following, we will discuss a COT treatment of two-electron atoms
explicitly including the triple collision dynamics when approaching
the limit $E\to 0_-$.  Introducing the hyperradius $R = ({\bf r}^2_1 + 
{\bf r}^2_2)^{1/2}$, we fix $R_0$
such that a surface $\Sigma$
defined as $R = R_0$ encloses the support of initial state $\phi_i$.
The surface $\Sigma$ naturally leads to a partition
of the configuration space into physically distinct regions. 
In particular, quantum contributions to (\ref{cs-G}) from the 
inner region are insensitive to the total energy. Contributions
from the outer region test the full scale of the classically allowed region
of size $|E|^{-1}$ and will be responsible for the resonance structures
near the double ionisation threshold $E=0$.
Following Granger and Greene \cite{GG00}, we write the photo-ionisation
cross section (\ref{cs-G}) in terms of local scattering matrices, that is, 
\begin{eqnarray} \label{cs-S1}
\sigma &=& 4 \pi^2 \alpha \hbar \omega \, \Re \left[\, d^\dagger
\left( 1 - S^{\uparrow}S^{\downarrow}\right)^{-1}
\left( 1+S^{\uparrow}S^{\downarrow}\right) d\right]\\ 
\label{cs-S2} &=&4 \pi^2 \alpha \hbar \omega \, \Re \left[\,
d^\dagger\left( 1 + 2 S^{\uparrow}S^{\downarrow} +
2 (S^{\uparrow}S^{\downarrow})^2 + \ldots \right) d \right] \, .
\end{eqnarray}
Here, $S^{\downarrow}$ is a core-region scattering matrix which maps 
amplitudes of waves coming in at $\Sigma$
onto amplitudes of the wave components 
going out at $\Sigma$; it thus contains
all the information about the correlated two-electron dynamics 
near the nucleus. Likewise, $S^{\uparrow}$ describes the wave dynamics
of  the two-electron wave function emanating from and returning to $\Sigma$
and thus picks up long-range correlation in the exterior of $\Sigma$. 
Furthermore, $d$ is the atomic dipole vector, 
$ d_n(E) = <\Psi_n^\downarrow(E) | D \phi_i > $
with $\Psi_n^\downarrow(E)$ being the $n$th linearly-independent
energy-normalised solution of the Schr\"odinger equation inside $\Sigma$
with incoming wave boundary conditions at $\Sigma$ \cite{GG00}.
This type of scattering formulation was independently developed
in \cite{Pro95} for general surfaces $\Sigma$,
for a semiclassical formulation, see \cite{Bog92}. The series
expansion (\ref{cs-S2}) was exploited in \cite{Del94,QD,series} in order
to include core-region scattering or quantum defect effects in COT.  

For the wave dynamics inside $\Sigma$, the double ionisation threshold 
$E=0$ is irrelevant and the core-region scattering matrix $S^{\downarrow}$
as well as the dipole vector $d$ vary smoothly across $E=0$; they 
can be regarded as constant for energies sufficiently close to the threshold.
The information about the increasing number of overlapping resonances near 
the threshold is thus largely contained in $S^{\uparrow}$.

Semiclassical approximations for the quantities introduced above
are valid in the outer-region $R > R_0$. The long-range scattering matrix,
$S^\uparrow$, can thus be treated semiclassically while $d$
and $S^\downarrow$ demand a full quantum treatment.
The semiclassical representation of $S^\uparrow$ in position space reads
\cite{GG00,Pro95,Bog92}
\begin{equation} \label{G-semi}
S^{\uparrow}(x,x',E)\approx (2\pi i\hbar)^{-\frac{f-1}{2}} \sum_j 
|M_{12}|^{-1/2}_j e^{i\frac{S_j}{\hbar}-i\frac{\pi\nu_j}{2}}\, ,
\end{equation}
where the sum is taken over all classical paths $j$ connecting
points $x$ and $x'$ on $\Sigma$ without crossing $\Sigma$;
$S_j(E)$ denotes the action of that path, $\nu_j$ is
the Maslov index and $f=4$ is the dimension of the system for
fixed angular momentum. Furthermore, 
$|M_{12}|^{-1/2}_j = |\partial^2 S_j(x, x')/ 
\partial x \partial x'|^{1/2}$, where 
$M_{12}$ is a $(3 \times 3)$ sub-matrix of the $6$-dim.\ 
Monodromy matrix  describing the linearised flow near a trajectory.
Note that due to the strong instability of the classical dynamics 
near the triple collision, these matrix elements become singular 
for triple collision orbits (TCO)
starting from or falling into the triple collision $R = 0$.
It is thus important here that trajectories 
contributing to (\ref{G-semi}) start at a fixed hyper-radius $R_0 >0$ 
away from the triple collision \cite{note1}.

Making use of the scaling properties of the classical dynamics, we introduce
the transformation \cite{RTW93}
\[ {\bf r} = \tilde{\bf r}/|E|; \quad {\bf p} = \sqrt{|E|}\tilde{\bf p};
\quad S = \tilde{S}/\sqrt{|E|}, \quad {\bf L} = \tilde{\bf L}/\sqrt{|E|}\]
where $\tilde{\bf r}, \tilde{\bf p}$ corresponds to coordinates and
momenta at fixed energy $E = -1$ and $\bf L$ is the total angular momentum.
Expressing $\bf L$ in scaled coordinates,
we have $\tilde{\bf L} \to 0$ as $E\to 0$ 
and can thus restrict the analysis to the three degrees of freedom 
subspace $\tilde{L} = 0$ (for fixed $L$) \cite{CLT04}. 

In scaled coordinates, the inner-region shrinks according to 
$\tilde{R}_0 = |E| R_0 \to 0$ for $E\to 0_-$, and
the part of the dynamics contributing to $S^\uparrow$ in 
(\ref{G-semi}) is formed by trajectories starting and ending closer and 
closer to the triple collision $R =0$ as  $|E| \propto \tilde{R}_0 \to 0$.
TCOs only occur in the so-called eZe space 
\cite{CLT04}, a collinear subspace of the full three body 
dynamics where the two electrons are on opposite sides of the nucleus 
\cite{RTW93}. As $\tilde{R}_0 \to 0$, only orbits coming close to the 
eZe space can start and return to $\tilde{\Sigma}$ and they will do so in 
the vicinity of a {\em closed triple collision orbit} (CTCO) starting 
and ending exactly in the triple collision. The dynamics in the eZe space 
is relatively simple as it is conjectured to be fully chaotic 
with a complete binary symbolic dynamics. In particular, for every finite 
binary symbols string there is a CTCO, the shortest being the so-called
Wannier orbit (WO) of symmetric collinear electron dynamics.
Furthermore TCOs escape from or approach the triple collision at 
$R=0$ always symmetrically along the $r_1 = r_2$ axis in the 
eZe space, that is, along the WO \cite{McG74}. This universality will 
be exploited below when treating the energy-dependence of $M_{12}$
in (\ref{G-semi}).

Returning to the cross-section (\ref{cs-S1}), we
write $\sigma = \sigma_0 + \sigma_{fl}$, where we identify the smooth
contribution $\sigma_0$ with the leading term in the series expansion
(\ref{cs-S2}). The main contribution to the fluctuating part of the cross section 
$\sigma_{fl}$ is contained in $S^\uparrow$ which in semiclassical 
approximation (\ref{G-semi}) can  be expressed in terms of orbits returning 
to $\tilde{\Sigma}$ once; multiple traversals of $\tilde{\Sigma}$ represented 
by $\left(S^\uparrow S^\downarrow \right)^k$ with $k \ge 2$ will give 
sub-leading contributions in the semiclassical limit $E \to 0_-$ due to 
the unstable dynamics near the triple collision. Furthermore, swarms of 
trajectories starting on $\tilde\Sigma$ and returning to $\tilde\Sigma$  
will do so close to the eZe subspace and thus in the neighbourhood of a 
CTCO with actions and amplitudes approaching those of the CTCO trajectory 
as $\tilde{R}_0 \to 0$. The fluctuating part can thus in leading order be
written in the form
\begin{equation} \label{s_fl_2}
\sigma_{fl}(E) \approx
\Re \sum_{{\rm CTCO}_j} A_j(E)
e^{i z \tilde{S}_j } 
\end{equation}
with
\begin{equation} \label{M_j}
A_j(E) \propto |M_{12}(E)|_j^{-1/2}
= |E|^{9/4} |\tilde{M}_{12}(\tilde{R}_0)|_j^{-1/2}
\end{equation}
and $z = 1/\hbar\sqrt{|E|}$.
In (\ref{s_fl_2}),
the sum is taken over all CTCO's $j$ starting and ending at 
$\tilde{\Sigma}$. Note that the stability $\tilde{M}_{12}(\tilde{R}_0)$ in 
scaled coordinates depends on energy implicitly through the scaled radius
$\tilde{R}_0 (E) = |E| R_0$. As $E\to 0_-$, $\tilde{M}_{12}$ picks up 
additional contributions of parts of the CTCO closer and closer to 
the triple collision. Asymptotically, all CTCOs approach the triple collision 
along the WO and the contributions to $M_{12}$ become
orbit-independent.
The $R$-dependence for the Monodromy matrix of the WO can for small $\tilde{R}$ 
be obtained {\em analytically} \cite{NNC06} leading to 
\begin{equation} \label{tildeM_j}
|\tilde{M}_{12} (\tilde{R}_0)| \propto |\tilde{R_0}|^{-2\mu + 9/2}
\; \mbox{for} \; \tilde{R}_0 \to 0
\end{equation}
with exponent
\begin{equation} \label{mu_t}
\mu= \mu_{eZe} + 2 \mu_{wr} =
\frac{1}{4}\left[\sqrt{\frac{100 Z-9}{4Z-1}} + 
2\sqrt{\frac{4 Z - 9}{4Z -1}}\right].
\end{equation}
Thus, in unscaled coordinates, $M_{12}$ in (\ref{M_j})
diverges which is a direct consequence 
of the non-regularisability of the triple collision acting as an infinitely 
unstable point in phase space; details will be presented in \cite{NNC06}.  
The exponent $\mu$ in (\ref{mu_t}) is universal for all CTCOs and 
consists of two components: $\mu_{eZe}$ is  related to the linearised dynamics 
in the eZe space and $\mu_{wr}$ picks up 
contributions from two equivalent expanding degrees of freedom orthogonal
to the eZe space in the so-called Wannier ridge (WR). The latter is 
the invariant subspace of symmetric electron dynamics with $|r_1| = |r_2|$ at 
all times \cite{RTW93}.  The fluctuations in the photo-ionisation cross 
section thus vanish in amplitude as $E \to 0_-$ according to 
\begin{equation} \label{CTCO1}
\sigma_{fl}(E) \propto |E|^{\mu} \, \Re \sum_{{\rm CTCO}_j} a_j
e^{i z \tilde{S}_j}
\end{equation}
where the rescaled amplitudes $a_j = |E|^{-\mu} A_j$ depend only weakly on $E$.
These 
amplitudes contain contributions from the linearised dynamics along the 
orbit far from $\Sigma$ as well as information about the inner quantum region 
$R<R_0$ via the dipole vector $d$ and the core-region scattering matrix
$S^\downarrow$.
Furthermore, multiple traversals of $\Sigma$ contained in
$\left(S^\uparrow S^\downarrow\right)^k$ with $k \ge 2$ in (\ref{cs-S2})
approach the triple collision $k$ times from the semiclassical 
side and will thus contribute at lower order 
with weights scaling at least as $A_{kj} \sim |E|^{k \mu}$. 
The exponent $\mu$
in (\ref{mu_t}) is different from Wannier's exponent $\mu_w$ with
\begin{equation}\label{mu_W}
\mu_w = \frac{1}{4}\sqrt{\frac{100 Z - 9}{4Z -1}} - \frac{1}{4}\, . 
\end{equation}
One obtains, for example, $\mu = 1.30589...$ compared to 
$\mu_w = 1.05589...$ for helium; the WR contributes to the decay 
for $Z> 9/4$ when $\mu_{wr}$ is real.

The exponent $\mu$ can be interpreted in terms of stability
exponents of the triple collision singularity.
Using an appropriate scaling of space and time by, for example, employing McGehee's 
technique \cite{McG74},
the dynamics near the singularity is dominated by 
two unstable fixed points in scaled phase space,
the {\em double escape point} (DEP) and the {\em triple collision point} (TCP).
In unscaled coordinates, these fixed 
points correspond to the WO at energy $E=0$, that is, the DEP is the 
trajectory of symmetric double escape while
the TCP corresponds to the symmetric triple collision and is the time reversed
of the DEP. The triple collision itself can be mapped onto the
classical dynamics at $E=0$; likewise, $\tilde{R}_0 = |E| R_0$ acts as 
a parameter measuring the closeness to the $E=0$ manifold which contains the
fixed points, see \cite{CLT04,McG74}. Most classical trajectories emerging from 
$\tilde{\Sigma}$ in the vicinity of the triple collision $\tilde{R} =0$ will lead to 
immediate ionisation of one electron carrying away a larger 
amount of kinetic energy. Only a fraction of orbits starting on $\tilde{\Sigma}$ 
near the WO will enter a chaotic scattering region; the WO at $E<0$ thus acts as 
an unstable direction of the DEP, $U_D^{wo}$, with a stability exponent
$\lambda_{U_D}^{wo}$. From there, they can return to 
the triple collision region and thus approach the surface $\tilde{\Sigma}$ again 
along the WO, that is, along the stable direction $S_T^{wo}$.
The transition from the DEP into the chaotic scattering region
and from this scattering region to the TCP is limited by the least stable
eigendirections of the fixed points in each of 
the invariant subspaces (eZe or WR) perpendicular to the WO.

Trajectories leaving $\tilde{\Sigma}$ along the WO in the eZe space diverge  
from the WO along an unstable direction $U_D^{eZe}$ with a stability exponent
$\lambda_{U_D}^{eZe}$.
Competition of the instability in $U_D^{wo}$ with that in $U_D^{eZe}$ 
determines the fraction $\Delta_{eZe}^D$ of orbits entering the chaotic 
scattering region.  By using methods as in \cite{CLT04,Sie41}, one finds that
\begin{equation} \label{DeltaD}
\Delta_{eZe}^D \propto \tilde{R}_0^{\lambda_{U_D}^{eZe}/\lambda_{U_D}^{wo}}
= \tilde{R}_0^{\mu_w}\, 
\end{equation}
with exponent equal to Wannier's exponent (\ref{mu_W}). For 
the cross section (\ref{cs-G}), information about the phase 
space region returning from the chaotic scattering region
to the surface $\tilde{\Sigma}$ along the WO is also needed.
While approaching $\tilde{\Sigma}$, these orbits are  
deflected away from the triple collision
along an unstable direction $U_T^{eZe}$ of the TCP fixed point.
The fraction of orbits reaching the surface $\tilde{\Sigma}$
among those leaving the chaotic scattering region
scales thus as in (\ref{DeltaD}) now with exponent 
$\lambda_{U_T}^{eZe}/|\lambda_{S_T}^{wo}|$ .
Similar mechanisms apply for the WR dynamics.

The fraction $\Delta$ of two-electron trajectories making the transition 
from $\tilde{\Sigma}$ back to $\tilde{\Sigma}$ can thus in the limit 
$E \to 0_-$ be estimated in terms of the stability exponents of the fixed 
points, also referred to as Siegel exponents \cite{Sie41} in celestial 
mechanics. One obtains $\Delta \propto \tilde{R}_0^{2\mu}$ with 
$\mu$ as in (\ref{mu_t}) which can be written as  
\[
\mu_{eZe} = \frac{1}{2} 
\left(\frac{\lambda_{U_D}^{eZe}}{\lambda_{U_D}^{wo}} +
\frac{\lambda_{U_T}^{eZe}}{\left|\lambda_{S_T}^{wo}\right|}\right); \;\;
\mu_{wr} = \frac{1}{2} \left(\frac{\lambda_{S_D}^{wr}}{\lambda_{U_D}^{wo}} +
\frac{\lambda_{U_T}^{wr}}{\left|\lambda_{S_T}^{wo}\right|}\right);
\]
for the actual values of the stability exponents $\lambda$, see
\cite{Wa53, CLT04}.  The mean amplitude of 
the fluctuations in the quantum signal is thus asymptotically related to the 
fraction of phase space volume starting and ending at $\tilde{R}_0$.\\ 
\begin{figure}
\includegraphics[scale=.32]{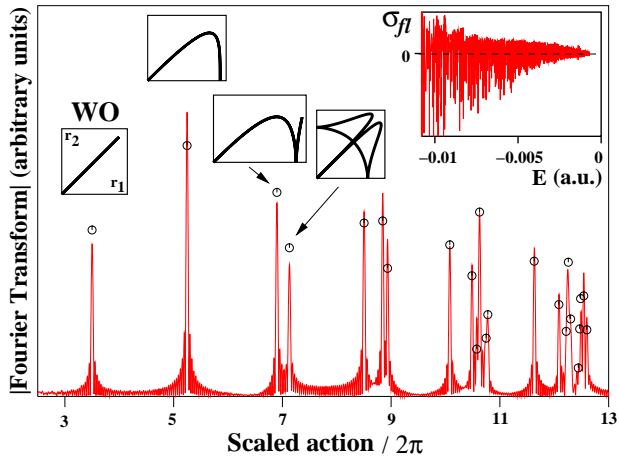}
\caption[]{\small The Fourier spectrum of the fluctuating part of the 
eZe cross section rescaled according to (\ref{Fz}); the circles denote
the position $\tilde{S}_j$ and (relative) size of $|M_{12}|^{-1/2}_j$ 
for CTCO's with $\tilde{S}/2 \pi < 13$. Corresponding trajectories 
in configuration space are shown for the first 4 peaks. 
Inset: $\sigma_{fl}$ for $N \le 52$.
}
\label{fig:fig1}
\end{figure}

A numerical study of the full three-body quantum problem is still out 
of reach for energies $E < I_N$ with $N \sim 15$ \cite{Pue01,Bue95};
we therefore chose a model system, namely eZe collinear helium 
first studied quantum mechanically in \cite{TH74}. 
We calculated the cross 
section (\ref{cs-G}) directly in a large set of basis functions using the 
method of complex rotation and obtained a converged signal for $N \sim 55$. 
Semiclassically, we consider now the dynamics in the eZe space alone, 
which contains all the important parts 
regarding the algebraic decay in the fluctuations.
The number of basis functions used are scaled with energy
to cover a fixed scaled region in $\tilde{R}$ containing CTCOs
with $\tilde{S}/2\pi \le 20$. 
Adopting the basis functions used by P\"uttner {\em et al} \cite{Pue01}
leading to a strongly banded Hamiltonian matrix, it is possible to increase the
basis size to $10^6$.
Starting with an odd initial state 
$\phi_i$, we obtain the cross section for the even parity eZe spectrum; 
its fluctuating part after numerically subtracting a 
smooth background is shown in Fig.\ \ref{fig:fig1}. The numerical value 
of the exponent $\mu$ is determined by rescaling the signal according to
\begin{equation} \label{Fz}
F(z) = |E|^{-\mu} \sigma_{fl}(z) / \hbar \omega
\end{equation}
and testing the stationarity of the Fourier transform of $F(z)$ in 
different energy windows \cite{NNC06}.  The best value thus obtained 
is $\mu = 1.306 \pm 0.035$ in good agreement with the
theoretical prediction (\ref{mu_t}). (Note that the real parts of
the exponents for 3-dim. helium and for eZe helium coincide
as $\Re \mu_{wr} = 0$). Furthermore, 
the peaks in the Fourier transform can be associated one-by-one with 
CTCO's in the eZe system, see Fig.\ \ref{fig:fig1}. We do not observe 
peaks associated with the concatenation of different CTCO's or repetitions 
of single CTCO's. This is consistent with the expected suppression of orbit 
contributions traversing $\Sigma$ more than once as discussed earlier. 
Furthermore, we calculated the geometrical contribution to the coefficients 
$a_j$ in (\ref{CTCO1}) directly from the matrix-elements $M_{12}$ by 
scaling out the leading order divergence; a clear correlation with the peak  
heights can be seen in Fig.\ \ref {fig:fig1}. Quantum 
contributions from the core region are thus indeed roughly the same for all
CTCOs.

In conclusion, we show that the fluctuations in the total photo-ionisation 
cross section below the double ionisation threshold follow an algebraic law 
with a novel exponent which can be
written in terms of  stability exponents of the triple collision. 
Our findings are verified numerically for a collinear model systems; we 
furthermore predict that the algebraic decay law is valid for the physically 
relevant 3 dimensional cases with an additional 
contribution from the WR dynamics for $Z>9/4$. Our findings will provide new 
impetus for experimentalists and theoreticians alike to study highly 
doubly excited states in two-electron atoms.

This work was supported in part by the Korea Research
Foundation (KRF-2006-521-C00019). We thank B.\ Gremaud, J.\ Madro\~{n}ero
and P.\ Schlagheck for useful discussions regarding the 
numerical calculations, and the KISTI supercomputing centre for numerical
support.


\begin{thebibliography}{999999}
\bibitem{note1} 
For regular potentials, Eqn.\ (\ref{G-semi}) can be written in terms of
closed orbits immediatly as contributions due to the finite radius become
negligible in the semiclassical limit.
\end{thebibliography}
\end{document}